\documentclass[%
reprint,
nofootinbib,
amsmath,amssymb,
aps,
prd,
]{revtex4-2}

\usepackage{multirow}
\usepackage{graphicx}
\usepackage{dcolumn}
\usepackage{bm}
\usepackage[bookmarks=true,bookmarksnumbered=true,colorlinks=true,urlcolor=black, citecolor=black,linkcolor=black,breaklinks=true]{hyperref}

\usepackage[table]{xcolor}
\usepackage{xcolor}
\usepackage{color}

\usepackage[T1]{fontenc}
\usepackage{braket}
\usepackage{xcolor}
\usepackage{tikz}
\usepackage{cleveref}

\newcommand{\vg}{``}

\newcommand{\Neff}{N_\mathrm{eff}}

\crefname{figure}{figure}{figures}
\Crefname{figure}{Figure}{Figures}

\definecolor{blue-violet}{rgb}{0.54, 0.17, 0.89}

\newcommand{\orcid}[1]{\begingroup
  \hypersetup{hidelinks}\href{https://orcid.org/#1}{\includegraphics[width=8pt]{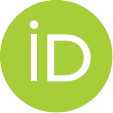}} \endgroup}

\begin{document}

\preprint{APS/123-QED}

\title{Probing the neutrino chemical potential with cosmological observations}

\begingroup
\renewcommand{\thefootnote}{\ensuremath{\alpha}}
\footnotetext[0]{\textit{These authors contributed equally to this work.}}
\endgroup

\author{Pietro Ghedini$^\alpha$ \orcid{0009-0001-4861-4867}}
\email{pietro.ghedini@ific.uv.es}
\affiliation{Instituto de Física Corpuscular (IFIC), CSIC‐UV}
\affiliation{Departament de Física Teòrica, UV, Spain}

\author{Riccardo Impavido$^\alpha$ \orcid{0009-0002-8711-4248}}
\email{mpvrcr@unife.it}
\affiliation{Dipartimento di Fisica e Scienze della Terra, Universit\`a degli Studi di Ferrara, via Giuseppe Saragat 1, 44122 Ferrara, Italy}
\affiliation{INFN, Sezione di Ferrara, via Giuseppe Saragat 1, 44122 Ferrara, Italy}

\author{Stefano Gariazzo \orcid{0000-0002-4160-6987}}
\email{gariazzo@ific.uv.es}
\affiliation{Instituto de F\'{i}sica Corpuscular (IFIC), CSIC‐Universitat de València, Spain}
\affiliation{University of Turin, Physics department and INFN, Sezione di Torino, Via P. Giuria 1, I–10125 Torino, Italy}

\author{Olga Mena \orcid{0000-0001-5225-975X}}
\email{omena@ific.uv.es}
\affiliation{Instituto de F\'{i}sica Corpuscular (IFIC), CSIC‐Universitat de València, Spain}

\author{Deng Wang \orcid{0000-0003-2062-5828}}
\email{dengwang@ific.uv.es}
\affiliation{Instituto de F\'{i}sica Corpuscular (IFIC), CSIC‐Universitat de València, Spain}

\date{\today}

\begin{abstract}
The electron neutrino degeneracy parameter, $\xi_{\nu_\mathrm{e}} = \mu_{\nu_\mathrm{e}} / T$, is tightly constrained by Big Bang Nucleosynthesis (BBN), while the degeneracy parameters of the other neutrino species, $\xi_{\nu_\mathrm{x}}$, remain weakly constrained by cosmological observations alone. In this manuscript we shall compute up-to-date bounds on $\xi_{\nu_\mathrm{e}}$ and $\xi_{\nu_\mathrm{x}}$ assuming that either they are constant free-parameters along the cosmic history or that they are redshift dependent quantities. In the latter case we employ a model-independent reconstruction approach based on the Piecewise Cubic Hermite Interpolating Polynomial (PCHIP) formalism with four nodes, located at $z\simeq$ 10, 100, 1000 and $10^8$. We shall also consider two scenarios for neutrinos, specifically three degenerate neutrinos ($\xi_{\nu_\mathrm{e}}$ = $\xi_{\nu_\mathrm{x}}$) and the case in which we actually differentiate between $\xi_{\nu_\mathrm{e}}$ and $\xi_{\nu_\mathrm{x}}$. We perform a cosmological analysis combining CMB data from Planck, SPT, and ACT with BAO measurements from DESI, showing the impact of including BBN observables from either EMPRESS results, which allow for a non-zero chemical potential, or from LBT observations, compatible with the standard $\xi_\nu$ = 0 prediction. We explicitly show that the BBN data, via the change in neutron-to-proton interconversion rates, mostly constrain $\xi_{\nu_\mathrm{e}}$, parameter for which we observe a preferred non-zero positive value at $95\%$~C.L. in the non-degenerate neutrino case at the BBN period. Since the Hubble constant is correlated with $\xi_{\nu}$, through $N_{\rm eff}$, a larger value of $H_0$ is allowed within these models, making them really interesting scenarios where to test non-standard physics models.
\end{abstract}

\maketitle

\section{\label{sec:introduction}Introduction}
Among the different issues cosmology is facing nowadays, one of the longstanding puzzles is the so-called Baryon Asymmetry of the Universe (BAU), the observed excess of matter over antimatter. This asymmetry is quantified via the baryon-to-photon ratio~\cite{Planck:2018vyg}

\begin{equation}
    \eta_\mathrm{B} = \frac{n_\mathrm{B} - n_{\bar{\mathrm{B}}}}{n_\gamma} \simeq 6.2 \times 10^{-10}~,
\end{equation}
\noindent which value is robustly constrained from BBN and CMB observations. 

On the other hand, the lepton asymmetries $\eta_{\mathrm{L},\,\alpha}$ remain less understood, even if they could have important impacts on the early Universe physics. Moreover, a non-zero (total) lepton asymmetry would affect the expansion rate of the Universe and the primordial Helium abundance. In principle, $\eta_\mathrm{L}$ could be order of magnitudes higher than the baryon one and it could be realised by having an excess of neutrinos over antineutrinos (or vice versa). One can generically parametrize the lepton asymmetry, neglecting the contribution of charged leptons\footnote{Due to charge neutrality, we expect the contribution of charged leptons to be of the same order as the baryon asymmetry, $\eta_\mathrm{B}$.}, in terms of the (dimensionless) neutrino chemical potential $\xi_\nu = \mu_\nu / T$, also known as the degeneracy parameter\footnote{Note that in the following we will actually place constraints and study the effect of the degeneracy parameter $\xi_\nu$, referring to it as the chemical potential.}~\cite{Iocco:2008va}

\begin{equation}
\begin{split}
    \eta_\nu &= \frac{1}{n_\gamma}\sum_{\alpha = e, \mu, \tau}(n_{\nu_\alpha}-n_{\bar{\nu}_\alpha})\\&=\frac{1}{12\zeta(3)}\sum_{\alpha = e, \mu, \tau}\left[\frac{T_{\nu_\alpha}}{T_\gamma}\right]^3\left(\pi^2\xi_{\nu_\alpha}+\xi_{\nu_\alpha}^3\right)~.
\end{split}
\end{equation}

The electron neutrino chemical potential is the one with the largest impact, since an excess of electron neutrinos or antineutrinos in the electroweak processes during BBN will alter the interaction rates ($n \leftrightarrow p$), impacting the formation of Helium and Deuterium and leaving observable imprints on the CMB temperature and polarization power spectra.
On the other hand, the muon and tau neutrino chemical potentials are much less constrained by BBN measurements because they do not directly participate in the weak interactions with primordial nucleons and therefore cosmological observations, primarily from the CMB (e.g., Planck) but also indirectly through a modified expansion history from BAO (e.g., DESI), allow significantly larger chemical potentials, with current bounds typically at the level of O(1).

allow these flavours to have larger values, typically constrained to \(\mathcal{O}(1)\). 
Over the past years, there have been different cosmological analyses trying to study the properties of the neutrino distribution function~\cite{deSalas:2018idd,Barenboim:2025vrc,Dai:2025pea,Alvey:2021sji}, also focusing on the chemical potential of neutrinos. Constraints have been derived from either CMB data alone~\cite{Oldengott:2017tzj,Barenboim:2016lxv,Nunes:2017xon,Bonilla:2018nau} or in combination with BBN measurements~\cite{Domcke:2025lzg,Domcke:2025jiy,Li:2024gzf}. The impact of a non-zero chemical potential on the cosmological tensions has also been explored in the literature~\cite{Li:2025rjr,Seto:2021tad}. While CMB data can probe the integrated effects of neutrino asymmetry, both on the evolution history and on the anisotropy power spectra, it is only when BBN is properly taken into account that we can derive tight constraints on the parameters of interest.

In this context, the primordial Helium abundance $Y_\mathrm{He}$ serves as a key quantity that connects BBN and CMB constraints, and its precise determination is necessary to properly study the implications of a non-zero chemical potential. The LBT collaboration has recently published a new measurement of $Y_\mathrm{He}$~\cite{Aver:2026dxv} in agreement with its standard prediction. Being the most precise measurement in the literature, it further increases the tension with the EMPRESS result~\cite{Matsumoto:2022tlr}, providing a value lower than the Standard Model one at the $3\sigma$ level. Allowing for a lower value of the Helium abundance opens the possibility of having a non-zero (electron) neutrino chemical potential, as explored in recent works~\cite{Escudero:2022okz,Li:2025rjr}. In particular, in~\cite{Li:2025rjr} it was shown that allowing for a non-zero value of the electron neutrino chemical potential, based on the EMPRESS results, could lead to an alleviation of the long-lasting $H_0$ tension.

The aim of this work is to provide state-of-the-art bounds on the possibility of having non-zero neutrino asymmetries in our Universe, including models with time-varying $\xi_\nu$. In particular, we will be using a model-independent reconstruction based on the Piecewise Cubic Hermite Interpolating Polynomial (PCHIP) formalism, allowing $\xi_\nu$ to phenomenologically vary with redshift, and probing therefore the origin of cosmological bounds.

This manuscript is organized as follows. \Cref{sec:pheno} describes the effects of the chemical potential on both CMB and BBN, also showing how the cosmological spectra are affected, specifically for the case of a time-dependent neutrino chemical potential. \Cref{sec:methodology} describes the reconstruction method and the cosmological scenarios we studied, together with the cosmological datasets and the adopted inference methodology. In \Cref{sec:results} we show and discuss our results. Finally, we present our conclusions and final remarks in \Cref{sec:conclusions}. 

\section{\label{sec:pheno} Phenomenology of the neutrino chemical potential}

\subsection{\label{sec:nu_chem} Effects on BBN and CMB}

\begin{figure*}
\centering
\includegraphics[width=\linewidth]{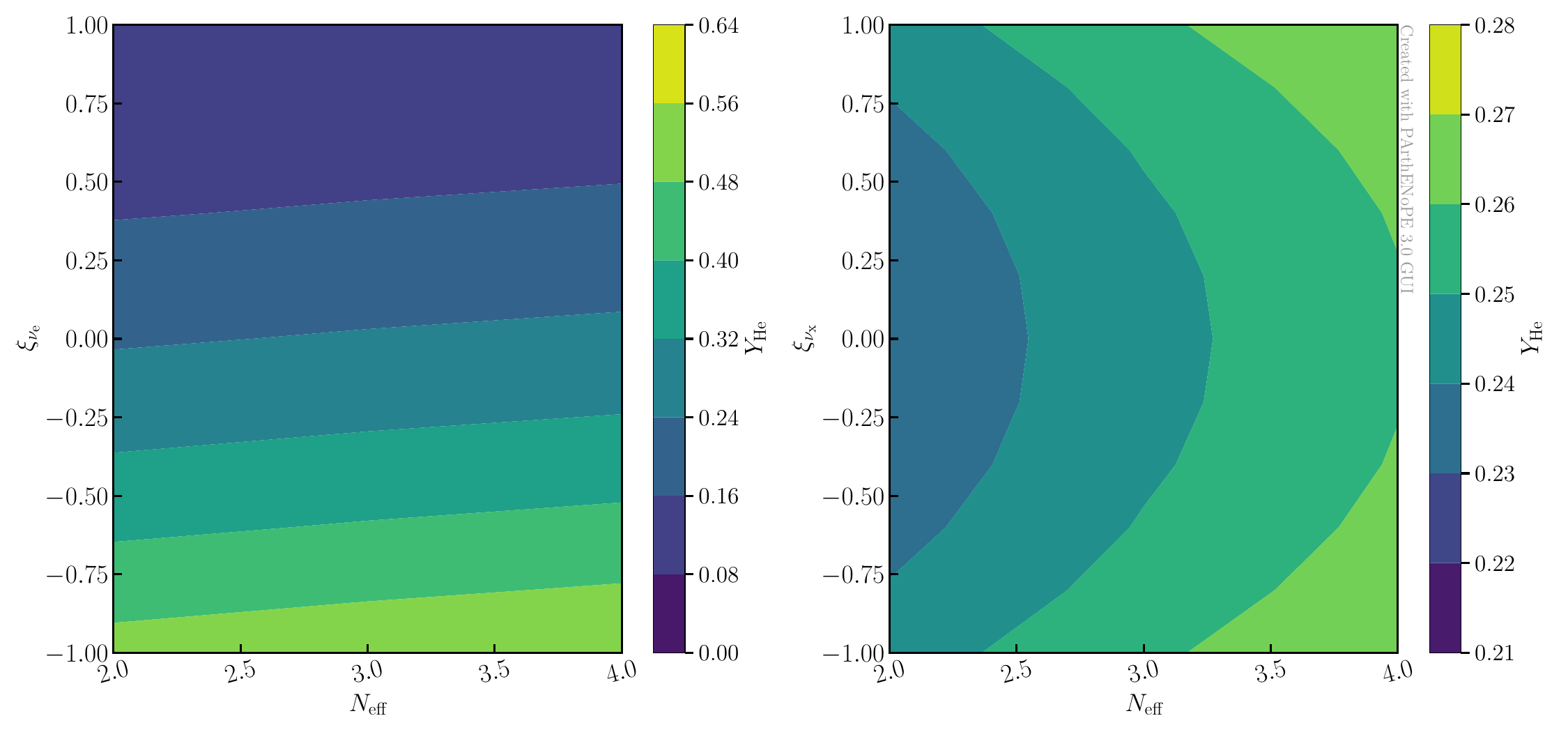}
\caption{Plot obtained with \texttt{PArthENoPE 3.0 GUI}~\cite{Gariazzo:2021iiu} showing the impact of the two different chemical potentials, $\xi_{\nu_\mathrm{e}}$ and $\xi_{\nu_\mathrm{x}}$, on $N_\mathrm{eff}$ and $Y_\mathrm{He}$. To obtain the left plot, we fix $\xi_{\nu_\mathrm{x}}$=0, while for the right one we fix $\xi_{\nu_\mathrm{e}}$=0. In both cases, we choose a value of $\eta_{10}=10^{10}\eta$, with $\eta$ the baryon-to-photon ratio, from our grid of points compatible at 1$\sigma$ with the standard BBN prediction, i.e. $\eta_{10} =  6.040 \pm 0.118$~\cite{ParticleDataGroup:2024cfk}.}\label{fig:parthenope}
\end{figure*}

The neutrino chemical potential has two main effects on cosmology, specifically on BBN and on the neutrino energy density, ofa different strength depending on the neutrino flavour. The BBN effect is mostly due to the fact that the electron neutrino chemical potential, $\xi_{\nu_\mathrm{e}}$, modifies the neutron-to-proton conversion rates during BBN~\cite{Steigman:2012ve}, shifting the neutron–proton equilibrium ratio at freeze-out and thereby increasing (decreasing), for negative (positive) $\xi_{\nu_\mathrm{e}}$, the primordial Helium-4 abundance, expressed by the fraction $Y_\mathrm{He}$~\cite{Pitrou_2018}

\begin{equation}
    Y_\mathrm{He} \simeq Y_\mathrm{He}^{\mathrm{sBBN}}e^{-0.96 \xi_{\nu_\mathrm{e}}}~,
\end{equation} 

\noindent where $Y_\mathrm{He}^{\mathrm{sBBN}} \equiv 0.24709 \pm 0.00017$~\cite{Pitrou:2018cgg} is $Y_\mathrm{He}$ in the standard BBN (sBBN) scenario. In \Cref{fig:parthenope}, obtained from the BBN code \texttt{PArthENoPE} \cite{Gariazzo:2021iiu}, we show the contour plots of $Y_\mathrm{He}$ and $N_\mathrm{eff}$ after varying the electron neutrino chemical potential $\xi_{\nu_\mathrm{e}}$ or the chemical potential of the other species $\xi_{\nu_\mathrm{x}}$. It is clear that $Y_\mathrm{He}$ is sensitive to the sign of $\xi_{\nu_\mathrm{e}}$ (left plot), while being also sensitive to the expansion history. However, $Y_\mathrm{He}$ is only affected by the absolute value of $\xi_{\nu_\mathrm{x}}$, which modifies the expansion history through $\Delta N_\mathrm{eff}$ via even powers of $\xi_\nu$ (see below).

Moreover, all chemical potentials $\xi_\nu$ enter the distribution function of cosmological neutrinos, expressed (for neutrinos and antineutrinos) as 

\begin{equation}
    f_0(q) = \frac{1}{e^{y-\xi_\nu}+1}+\frac{1}{e^{y+\xi_\nu}+1}~,
\end{equation}
where $y = q/ T_\nu$ is the dimensionless comoving momentum.
This effect changes the neutrino energy density, modifying the background expansion history at all times.
In the early Universe this affects the effective number of neutrinos $N_\mathrm{eff}$, 

\begin{equation}
\begin{split}
    \Delta N_\mathrm{eff}(\xi_\nu) & = N_\mathrm{eff}^\mathrm{tot} - N_\mathrm{eff}^\mathrm{std} \\ & \simeq \sum_{\alpha = e, \mu, \tau} \left[ \frac{30}{7}\left(\frac{\xi_{\nu_\alpha}}{\pi}\right)^2+ \frac{15}{7}\left( \frac{\xi_{\nu_\alpha}}\pi \right)^4\right]~,
\label{eq:deltaneff}
\end{split}
\end{equation}

\noindent where $\alpha$ runs over the species of neutrinos, affecting the primary CMB and, indirectly, BBN, and we assume $N_\mathrm{eff}^\mathrm{std}=3.044$. 

At late times, the change in $\Omega_\nu$, determined by the change in the neutrino number density $n_\nu$, reads as

\begin{equation}
    \Delta \Omega_\nu (\xi_\nu) = \sum_{\alpha = e, \mu, \tau} m_{\nu_\alpha} \left[\frac{2 \log(2)\,\xi_{\nu_\alpha}^2}{3\,\zeta(3)} + \frac{\xi_{\nu_\alpha}^4}{72\,\zeta(3)}\right]~,
    \label{eq:omeganu}
\end{equation} 

\noindent where $\alpha$ runs over the species of massive neutrinos with mass $m_{\nu_\alpha}$. This affects secondary CMB and the matter power spectrum through neutrino free-streaming. All effects on the background are only depending on even powers of the chemical potential, and thus cosmology alone cannot distinguish its sign. Beyond the changes in the background quantities, a change in the neutrino distribution function affects all integrated quantities in the Boltzmann hierarchy, modifying also the perturbation effects. 

\subsection{Effects on the cosmological spectra}
\label{sec:spectra}

\begin{figure*}
    \centering
    \includegraphics[width=1\linewidth]{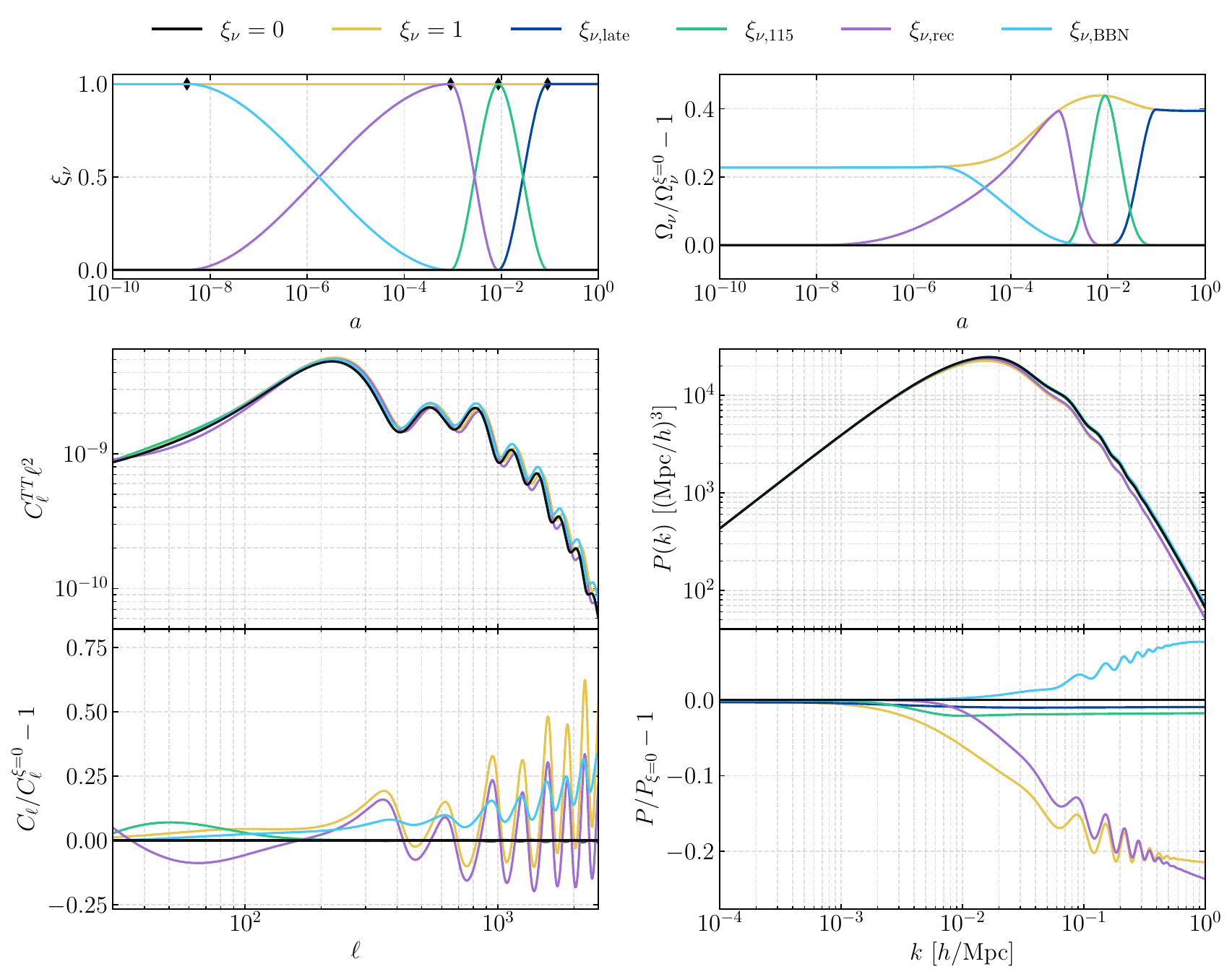}
    \caption{Phenomenological effect on linear spectra, all plots share the same colour code showed in the legend and all residuals are computed with respect to the scenario in which we have a constant $\xi_\nu=0$. \textit{Top-left:} PCHIP interpolations for various $\xi_\nu(z)$ scenarios for three massive (degenerate) neutrinos. Black diamonds are the PCHIP nodes. \textit{Top-right:} residual differences of energy density evolution. \textit{Bottom-left:} CMB linear temperature power spectrum $C_\ell^{TT}$ for a Universe with three degenerate massive neutrinos with $m_\nu = 0.02$ eV, so that $\sum m_\nu = 0.06$ eV, having a redshift dependent chemical potential. \textit{Bottom-right:} linear matter power spectrum. }
    \label{fig:pheno}
\end{figure*}

To illustrate the effects of a time varying chemical potential on the CMB and on the matter power spectrum, we refer to \Cref{fig:pheno}. We focus on three degenerate massive neutrinos and explore four models with time-varying chemical potential. Specifically, at four nodes in redshift (named $\xi_{\nu, \,\mathrm{late}}, \, \xi_{\nu, \, 115}, \, \xi_{\nu, \,\mathrm{rec}}, \, \xi_{\nu, \,\mathrm{BBN}}$, located respectively at $z=10, \, 115, \, 1100$ and $3\times 10^8$; see \Cref{tab:nodes} in \Cref{sec:PCHIP}), $\xi_\nu(z)$ transits from 0 to 1, and then to 0 again. We compare these cases to those in  which the chemical potential is constant, either 0 or 1, representing the values $\xi_\nu$ can assume in our (physical) limiting cases. 

As we are not compensating any of the effects that an increase of $\xi_\nu$, translated in an increase of $N_\mathrm{eff}$ or $\Omega_\nu$, have on the expansion history, the models $\xi_{\nu, \,\mathrm{rec}}$  and $\xi_\nu = 1$ (purple and yellow curves, respectively) undergo matter-radiation equality at redshift $z_\mathrm{eq}$ around $20\%$ smaller with respect to the rest of the models, due to the change to the energy density in relativistic matter around recombination. The change in the evolution of the neutrino energy density due to a time-depending chemical potential can be seen in the \textit{top-right} panel of \Cref{fig:pheno}. Its effects on CMB $C_\ell^{TT}$ linear temperature power spectrum and the linear matter power spectrum $P(k)$ are, instead, displayed in the lower panels of the same Figure. Therein, we can appreciate how changes at small scales (large $\ell $ or $k$) are present if $\xi_\nu(z)$ deviates from 0 at early times, while if $\xi_\nu$ differs from 0 at late times, effects on large scales are present.

\textbf{\emph{Matter power spectrum $\,-\,$}} In the \textit{bottom-right} panel of \Cref{fig:pheno} we show the impact of a time-varying neutrino chemical potential on the matter power spectrum. In the $\xi_{\nu, \,\mathrm{late}}$ model (blue curve) we observe a suppression at large scales due to an increased Hubble parameter at late times, while on small scales, where free-streaming is present, it displays an increased step-like suppression due to an increased $\Omega_\nu$, an effect similar to the one in the  $\xi_{\nu, \, 115}$ case (green curve). As anticipated, the models $\xi_{\nu, \,\mathrm{rec}}$ and $\xi_\nu = 1$ undergo matter-radiation equality at a later time, and this pushes the peak of the matter power spectrum to larger scales, explaining the suppression at small scales. Finally, the effect of $\xi_{\nu, \,\mathrm{BBN}}$ (light-blue curve) on $P(k)$ amounts to an increase of $N_\mathrm{eff}$ in the early times while keeping other quantities (namely the matter radiation equality and the baryon and cold dark matter abundances) fixed. This translates into an increase of $P(k)$ for scales smaller than the matter-radiation equality one.

\textbf{\emph{CMB temperature power spectrum $\,-\,$}} Regarding the $C_\ell^{TT}$ (\textit{bottom-left} panel of \Cref{fig:pheno}), the slight deviation of $\xi_{\nu, \,\mathrm{late}}$ is too small to be visible at this scale. The deviation of $\xi_{\nu, \, 115}$, instead, is due to a larger early ISW with respect to $\xi_\nu = 0$ (black curve); contrary to the standard late ISW, the effect is not a tangible enhancement of the first peak, but a slight enhancement of power on larger scales. The effect of $\xi_{\nu, \,\mathrm{rec}}$ is due to the different $z_\mathrm{eq}$ in the model, shifting the angular diameter distance at decoupling and thus changing the position of the first peak and the envelope of all the secondary peaks, analogously to what an increased $N_\mathrm{eff}$, without compensating anything else in the expansion history, would generate. The modifications generated by the transition of $\xi_{\nu, \,\mathrm{BBN}}$ are purely induced by the (anti-)correlation between the primordial Helium fraction $Y_\mathrm{He}$ and $\xi_\nu$. This is the only effect which is sensitive to the sign of the chemical potential. As a matter of fact, increasing $\xi_\nu$ at BBN implies reducing $Y_\mathrm{He}$, which increases the CMB temperature power in the damping tail as fewer baryons are bound in Helium, leaving more free electrons before recombination. This shortens the photon mean free path, leading to a reduction of the diffusion damping. 

In both spectra, we can see a combination of all the effects described above for the intermediate cases for the $\xi_\nu=1$ model (yellow curve). 

\section{\label{sec:methodology} Methodology and datasets}

\subsection{PCHIP formalism}
\label{sec:PCHIP}

One of the best methods to perform an agnostic data-driven determination of a particular function is via the Piecewise Cubic Hermite Interpolating Polynomial (PCHIP)~\cite{doi:10.1137/0905021, doi:10.1137/0717021}, which can preserve the shape of data distribution due to its special mathematical properties and avoid many of the oscillations and overshoots that ordinary cubic splines can produce. Overall, this method consists of a combination of third order polynomials that ensures smoothness between a set of input nodes. For more details on the PCHIP interpolation method, see for example the recent works of Refs.~\cite{Gariazzo:2014dla, Ghedini:2025epp}.

In this work, we apply this method to reconstruct the neutrino chemical potential, $\xi_\nu$, in a model-independent way, using four nodes at fixed redshifts. In this way, the functional form is completely determined once the values of the chemical potential at the four redshift values of our choice are specified.

The choice to work with only four nodes is dictated by the result of a previous simulation done with a larger (seven) set of nodes and by computational time: a higher number of nodes would translate into a higher number of parameters to be sampled in the MCMC. For the test simulation, we considered seven nodes at redshifts $z=0,\ 1,\ 10,\ 115,\ 1100,\ 3\times10^8,\ 10^{14}$. Given that data are not constraining the late time nodes (i.e. $z=0,\ 1,\ 10$), as the effects of the neutrino distribution function are weaker at such epochs, we decided to collapse~\footnote{A clarification is in order: with \vg collapse''  we mean that we still have an interpolation on seven nodes but we impose that the variables of interest in the "collapsed" nodes assume the same value.} them into a single node, specifically considering the one at $z=10$. We also collapse the last two nodes, $z=3\times10^8,\ 10^{14}$ into a single one to avoid time-varying $\xi_\nu$ during BBN.
In other words, in our PCHIP runs $\xi_\nu$ is constant below $z=10$ and above $z=3\times10^8$, respectively with values $\xi_{\nu, \, \mathrm{late}}$ and $\xi_{\nu, \, \mathrm{BBN}}$.

Therefore, as previously anticipated, we keep the nodes at redshift $z=10, 115, \, 1100, \, 3\times10^8$ to take into account the epochs at which a non-zero chemical potential is expected to have a more significant impact. Respectively, these nodes represent the epoch in which structure formation becomes non-linear, the epoch in which a neutrino with mass of around $0.06$ eV transitions from being relativistic to non-relativistic, the epoch of recombination and the time of BBN. The full set of nodes considered is shown in \Cref{tab:nodes}.

\begin{table}[h!]
\setlength{\tabcolsep}{5pt}
\renewcommand{\arraystretch}{1.4}
\centering
\begin{tabular} {c c c}
\textbf{Redshift} \boldmath{$z$} & \textbf{Scale factor} \boldmath{$a$} & \textbf{Parameter} \\
\hline
10 & $\approx$ 0.09 & $\xi_{\nu, \, \mathrm{late}}$\\
115 & $\approx$ 0.0086 & $\xi_{\nu, \, 115}$\\
1100 & $\approx$ $9\times10^{-4}$ & $\xi_{\nu, \, \mathrm{rec}}$\\
3$\times 10 ^8$ & $\approx$ $3\times10^{-9}$ & $\xi_{\nu, \, \mathrm{BBN}}$\\
\hline
\end{tabular}
\caption{Table summarizing the nodes chosen for the reconstruction. We show both the redshift and the equivalent scale factor, related by $1+z=a^{-1}$, specifying the nomenclature of the parameters that we sample.} 
\label{tab:nodes}
\end{table}

\subsection{Cosmological scenarios}
\label{sec:cosmo_scenarios}

We shall always assume a spatially flat Universe with adiabatic initial conditions. We consider two different scenarios, regarding how we treat neutrinos:

\textbf{\emph{Case $\boldsymbol{3 \, \nu}$}$\,-\,$} We consider three degenerate massive neutrinos, meaning that we are assuming $\xi_{\nu_\mathrm{e}} = \xi_{\nu_\mu} = \xi_{\nu_\tau} \equiv\xi_{\nu}$. We shall also fix the sum of the neutrino masses to be $\sum m_\nu = 0.06$ eV, which is the lowest limit allowed by neutrino oscillation experiments~\cite{Esteban:2024eli,deSalas:2020pgw,Capozzi:2021fjo}.

\textbf{\emph{Case $\boldsymbol{(1 + 2) \, \nu}$}$\,-\,$} We consider a very light neutrino state (m$_\mathrm{lightest} = 10^{-5}$ eV), that can be connected to the electron neutrino, as it would be the case for the normal hierarchy, with chemical potential $\xi_{\nu_\mathrm{e}}$. We then consider the remaining two neutrinos to be degenerate, massive ($\sum m_{\nu, \, \mathrm{degenerate}} = 0.06$ eV) and with the same chemical potential, $\xi_{\nu_\mu} = \xi_{\nu_\tau} \equiv \xi_{\nu_\mathrm{x}}$~\footnote{The names of the two chemical potential are given consistently with the nomenclature used in \texttt{PArthENoPE}.}.

In principle, each neutrino family could have a different chemical potential. The effect of neutrino oscillations, however, tends to re-equilibrate the difference between the three neutrino families~\cite{Froustey:2021azz} even when the initial values are quite apart from one another.
In this sense, it is perfectly reasonable to assume that at least two neutrino families share the same degeneracy parameter.

We introduce an additional classification, regarding how we treat the neutrino chemical potential $\xi_\nu$. Specifically, we consider the two following scenarios:

\textbf{\emph{Case $\boldsymbol{\xi_{\nu,\,\mathrm{constant}}}$}$\,-\,$} In this case, we simply consider the neutrino chemical potential to be constant over time, leaving it as a free parameter in the MCMC analyses.

\textbf{\emph{Case $\boldsymbol{\xi_{\nu,\,\mathrm{PCHIP}}}$}$\,-\,$} In this scenario, instead, we reconstruct the chemical potential using the PCHIP interpolation method with four nodes, as described in the previous section.\footnote{Note that in the $\boldsymbol{\xi_{(1 + 2)\nu}}$ case, we reconstruct both $\xi_{\nu_\mathrm{e}}$ and $\xi_{\nu_\mathrm{x}}$.} The aim of this reconstruction is not to set bounds on $\xi_\nu$ at different epochs, but to prove that BBN is the most sensitive epoch to $\xi_\nu$, testing the origin of the bounds one usually derives on the chemical potential by performing a cosmological analysis.

In the following we shall consider all the possible combinations of the four scenarios described above, namely we consider the cases \textbf{\emph{$\boldsymbol{\xi_{3\nu,\,\mathrm{constant}}}$}}, \textbf{\emph{$\boldsymbol{\xi_{3\nu,\,\mathrm{PCHIP}}}$}}, \textbf{\emph{$\boldsymbol{\xi_{(1 + 2)\nu,\,\mathrm{constant}}}$}} and \textbf{\emph{$\boldsymbol{\xi_{(1 + 2)\nu,\,\mathrm{PCHIP}}}$}}.

\subsection{Inference and measurements}
\label{sec:cosmo_analysis}

To constrain the parameters of our models, we use a modified version of the Cosmic Linear Anisotropy Solving System code (\texttt{CLASS})\footnote{The modified version of \texttt{CLASS} that supports the findings of this article is not publicly available, but it can be obtained from the authors upon reasonable request.}~\cite{Lesgourgues:2011re, Blas:2011rf}. The implemented modifications aim to compute the neutrino distribution function for the case of a non-zero chemical potential, to include the PCHIP reconstruction method and to modify the interpolation performed on the BBN tables. Specifically, the BBN tables currently in \texttt{CLASS} do not account for a non-zero neutrino chemical potential. To obtain the new table, we used the BBN code \texttt{PArthENoPE}~\cite{Pisanti:2007hk, Consiglio:2017pot, Gariazzo:2021iiu}. We then performed the inference of the cosmological parameters using \texttt{Cobaya}~\cite{Torrado:2020dgo, 2019ascl.soft10019T} with the convergence set as a Gelman-Rubin test~\cite{Gelman:1992zz} of $R-1.$ For the cases $\xi_{\nu,\mathrm{constant}} $ our aim is to place limits, and therefore we require a  convergence  of $R-1 \leq 0.01$, while for the scenarios $\xi_{\nu, \mathrm{PCHIP}}$ where our aim is to understand the source of the constraining power, we relax the threshold at $R-1 \leq 0.04$. The posterior distributions are obtained and analyzed using the \texttt{GetDist} package~\cite{Lewis:2019xzd}. 

Finally, we perform a Bayesian model comparison. For each model $\mathcal{M}_i$, we compute the Bayesian evidence\footnote{We actually compute its average value, with the associated error, obtained considering different values for the parameters \texttt{burnlen}, controlling the burn-in region of the chains, and \texttt{thinlen}, controlling the thinning, thus the markovianity, of the chains.} using the \texttt{MCEvidence} package~\cite{Heavens:2017afc, Heavens:2017hkr}, defined as

\begin{equation}
    \mathcal{Z}_i = \int d\theta \mathcal{L}(d|\theta,\mathcal{M}_i)\pi(\theta|\mathcal{M}_i)~.
\end{equation}

Specifically, we make use of a \texttt{Cobaya}-friendly version of \texttt{MCEvidence} as presented in the \texttt{wgcosmo} Github repository\footnote{\protect\url{https://github.com/williamgiare/wgcosmo/tree/main/statistics/MCMC_Evidence}}. We compare the models considered in this work with a $\Lambda$CDM-like model, where we consider $\xi_\nu=0$ but assume massive neutrinos, distinguishing between the $3\nu$ and $(1+2)\nu$ cases.

Model comparison is performed computing the logarithm of the Bayes factor
\begin{equation}
    \ln\mathcal{B}_{ij}=\ln\mathcal{Z}_i - \ln\mathcal{Z}_j~.
\end{equation}
A positive (negative) $\ln\mathcal{B}_{ij}$ means that our $\Lambda$CDM model $\mathcal{M}_i$ is favored (disfavored) over the specific model $\mathcal{M}_j$ considered in this work. Using the revised Jeffreys' scale~\cite{Kass:1995loi}, the evidence is classified as \textit{inconclusive}\footnote{This means that the two models perform comparably.} ($0\leq|\ln\mathcal{B}_{ij}|<1$), \textit{weak} ($1\leq|\ln\mathcal{B}_{ij}|<2.5$), \textit{moderate} ($2.5\leq|\ln\mathcal{B}_{ij}|<5$), \textit{strong} ($5\leq|\ln\mathcal{B}_{ij}|<10$) and \textit{very strong} ($|\ln\mathcal{B}_{ij}|\geq10$).

In all the analyses performed, we sample over the usual $\Lambda$CDM set of parameters, i.e. $\{ \omega_b,\, \omega_c,\, 100 \, \theta_s, \, \tau_{\rm reio}, \, \ln(10^{10}A_s), \, n_s \}$, to which we add the additional parameters characteristic of each cosmological scenario considered, as explained in \Cref{sec:cosmo_scenarios} and shown in \Cref{tab:priors}, together with the priors imposed on all the parameters. Note that all analyses are conducted with a fixed $N_\mathrm{eff}^\mathrm{std}=3.044$, where we actually mean that we keep fixed some of the \texttt{CLASS} parameters governing the neutrino sector, specifically \texttt{deg\_ncdm}, \texttt{N\_ncdm} and \texttt{N\_ur}. As can be seen from \Cref{eq:deltaneff}, the total $N_\mathrm{eff}$ will be given by $N^\mathrm{tot}_\mathrm{eff} = N_\mathrm{eff}^\mathrm{std} + \Delta N_\mathrm{eff}(\xi_\nu)$ and, in the \textbf{PCHIP} case it will not be a constant during cosmic evolution since the contribution given by $\xi_\nu$ will vary over redshift.

\begin{table}[h!]
\centering
\setlength{\tabcolsep}{6pt}
\renewcommand{\arraystretch}{1.35}
\centering
\begin{tabular}{cc}
\textbf{Parameters} & \textbf{Priors} \\
\hline
{\boldmath$\log(10^{10} A_\mathrm{s})$} & $\mathcal{U}[1.61, 3.91]$ \\
{\boldmath$n_\mathrm{s}$}              & $\mathcal{U}[0.8, 1.2]$ \\
{\boldmath$100\theta_\mathrm{s}$}      & $\mathcal{U}[0.5, 10]$ \\
{\boldmath$\Omega_\mathrm{b} h^2$}     & $\mathcal{U}[0.0074, 0.033]$ \\
{\boldmath$\Omega_\mathrm{c} h^2$}    & $\mathcal{U}[0.001, 0.99]$ \\
{\boldmath$\tau_\mathrm{reio}$}        & $\mathcal{U}[0.01, 0.8]$ \\
\hline
\hline
$\boldsymbol{\xi_\nu \, \, \mathrm{or} \, \, \xi_{\nu_\mathrm{e,\, x}}}$ & $\mathcal{U}[-1.0, 1.0]$ \\
$\boldsymbol{\xi_{\nu, \, \mathrm{BBN}}}$ & $\mathcal{U}[-1.0, 1.0]$ \\
$\boldsymbol{\xi_{\nu, \, i}}\, \, (i = \mathrm{late}, \, 115,\,\mathrm{rec})$ & $\mathcal{U}[-2.0, 2.0]$ \\
\hline
\end{tabular}
\caption{Uniform priors on the standard base cosmological parameters sampled in these analyses, plus the priors imposed on the neutrino chemical potential. Note that in the case of the PCHIP reconstruction, we impose a smaller prior on the node at BBN for consistency with how \texttt{PArthENoPE} works. For reasons related to the interpolation in \texttt{CLASS} and to the output of \texttt{PArthENoPE}, we also impose a more restrictive prior on $\Omega_\mathrm{b} h^2$. Note that the chemical potential priors on each node remain true also for $\xi_{\nu_\mathrm{e}}$ and $\xi_{\nu_\mathrm{x}}$
when they are both varying separately.}
\label{tab:priors}
\end{table}

We exploit the current state-of-the-art cosmological datasets to place constraints on the cosmological scenarios described above. In particular, we consider two different types of measurements as the baseline dataset:

\begin{itemize}
    \item \textbf{\emph{CMB measurements}}: we use CMB temperature data from the \texttt{Commander} likelihood~\cite{Planck:2018vyg,Planck:2019nip} and $E$-mode polarization data from \texttt{SRoll2}~\cite{Pagano:2019tci,Delouis:2019bub} at large angular scales ($2 \le \ell \le 29$), complemented at high multipoles by the \texttt{Plik\_lite} likelihood from the Planck mission. These are combined with the \texttt{ACT-lite}\footnote{\protect\url{https://github.com/ACTCollaboration/DR6-ACT-lite}} likelihood for ACT DR6~\cite{AtacamaCosmologyTelescope:2025blo} and the \texttt{SPT-lite}\footnote{\protect\url{https://github.com/SouthPoleTelescope/spt_candl_data}} likelihood for SPT-3G D1~\cite{SPT-3G:2025bzu, Balkenhol:2024sbv}. Following the prescriptions described in Refs.~\cite{AtacamaCosmologyTelescope:2025blo,SPT-3G:2025bzu}, we restrict ourselves to {\it Planck} data up to multipoles $\ell_{\rm max} = 1000$ in temperature and $\ell_{\rm max} = 600$ in polarization, and assume no correlations between SPT and the other datasets. 
    Furthermore, we include the joint \textit{Planck}+ACT DR6 lensing likelihood\footnote{\protect\url{https://github.com/ACTCollaboration/act_dr6_lenslike}}~\cite{ACT:2023kun,ACT:2023dou,ACT:2023ubw,Carron:2022eyg} and the MUSE lensing likelihood\footnote{\protect\url{https://github.com/qujia7/spt_act_likelihood}} for SPT-3G D1~\cite{SPT-3G:2024atg}. 
    In the following, we shall refer to this combined dataset combination as \textbf{SPA}. 
    \item {\textbf{\emph{BAO measurements}}}: we include the three-year data collection of Baryon Acoustic Oscillations (BAO) measurements from the Dark Energy Spectroscopic Instrument (DESI) presented in the second data release (DR2)~\cite{DESI:2025zgx}. We refer to this dataset as \textbf{DESI}.
\end{itemize}

Using the combination \textbf{SPA + DESI} as a baseline allows us to get constraining power from the early to the late Universe.

Since a non-zero chemical potential in the early Universe will impact BBN, we include two additional likelihoods based on Deuterium and Helium-4 abundances. Specifically, we compare the predictions on the Deuterium abundance obtained from \texttt{PArthENoPE} to the PDG value~\cite{ParticleDataGroup:2024cfk}, $D/H = ( 25.47 \pm 0.29)\times 10^{-6}$. On the other hand, we separately consider two different measurements of the Helium fraction, $Y_\mathrm{He}$, as listed in the following:

\begin{itemize}
    \item \textbf{\emph{EMPRESS}}: recent results by the EMPRESS survey~\cite{Matsumoto:2022tlr} report a value of the Helium fraction, $Y_\mathrm{He} = 0.2370^{+0.0034}_{-0.0033}$, which is $\sim 3\sigma$ lower than the standard BBN (sBBN) prediction, $Y_\mathrm{He} = 0.24709\pm 0.00017$~\cite{Pitrou:2018cgg}. As a consequence, a non-zero value for the electron neutrino chemical potential is derived by the collaboration, $\xi_{\nu_\mathrm{e}}=0.05^{+0.03}_{-0.02}$.
    \item {\textbf{\emph{LBT}}}: the LBT $Y_p$ project has recently reported a new measurement of the Helium fraction~\cite{Skillman:2026ltj,Aver:2026dxv,Yeh:2026pil}, which does not present a tension with the standard BBN predictions. The value measured by the LBT collaboration, $Y_\mathrm{He}=0.2458\pm0.0013$, is the most precise determination of the primordial $^4$He mass fraction to date.
\end{itemize}

We also take into account the theoretical errors introduced by the computations of \texttt{PArthENoPE}, as reported in~\cite{Gariazzo:2021iiu}, meaning that as reference (theoretical) values in the likelihood we will have $Y_\mathrm{He}$ = value $\pm \, \sigma \, \pm \, 0.00003 \, \pm 0.00012$ and $D/H$ = (value $\pm \, \sigma \, \pm \, 0.06 \, \pm 0.03)\times10^{-5}$.

Using the importance sampling method implemented in \texttt{Cobaya}, we re-evaluate our \textbf{SPA + DESI} chains to consider the constraining power of BBN on our models. 

\section{\label{sec:results} Results}

In this section we present the results of our analyses on a reduced set of parameters, showing $\xi_\nu$ in the case of $\xi_{\nu, \, \mathrm{constant}}$ or the nodes $\xi_{\nu, \, i}$ in the case of $\xi_{\nu, \, \mathrm{PCHIP}}$, together with some derived parameters, specifically the Hubble constant, $H_0$, the primordial Helium fraction, $Y_\mathrm{He}$, and either $N_\mathrm{eff, \, BBN}$~\footnote{$N_\mathrm{eff, \, BBN}$ refers to the value of $N_\mathrm{eff}$ computed at the BBN epoch by \texttt{CLASS} and used for the interpolation to compute $Y_\mathrm{He}$ from the tables obtained from \texttt{PArthENoPE}.} in the corner plots, or $\Delta N_\mathrm{eff, \, BBN}$, in the tables. We additionally show the Bayes factor $\ln\mathcal{B}_{ij}$ in the Tables.

\subsection{Case {\boldmath$\xi_{3\nu,\,\mathrm{constant}}$}}
\label{sec:3nuconst_results}

\begin{figure}[h!]
    \centering
    \includegraphics[width=\linewidth]{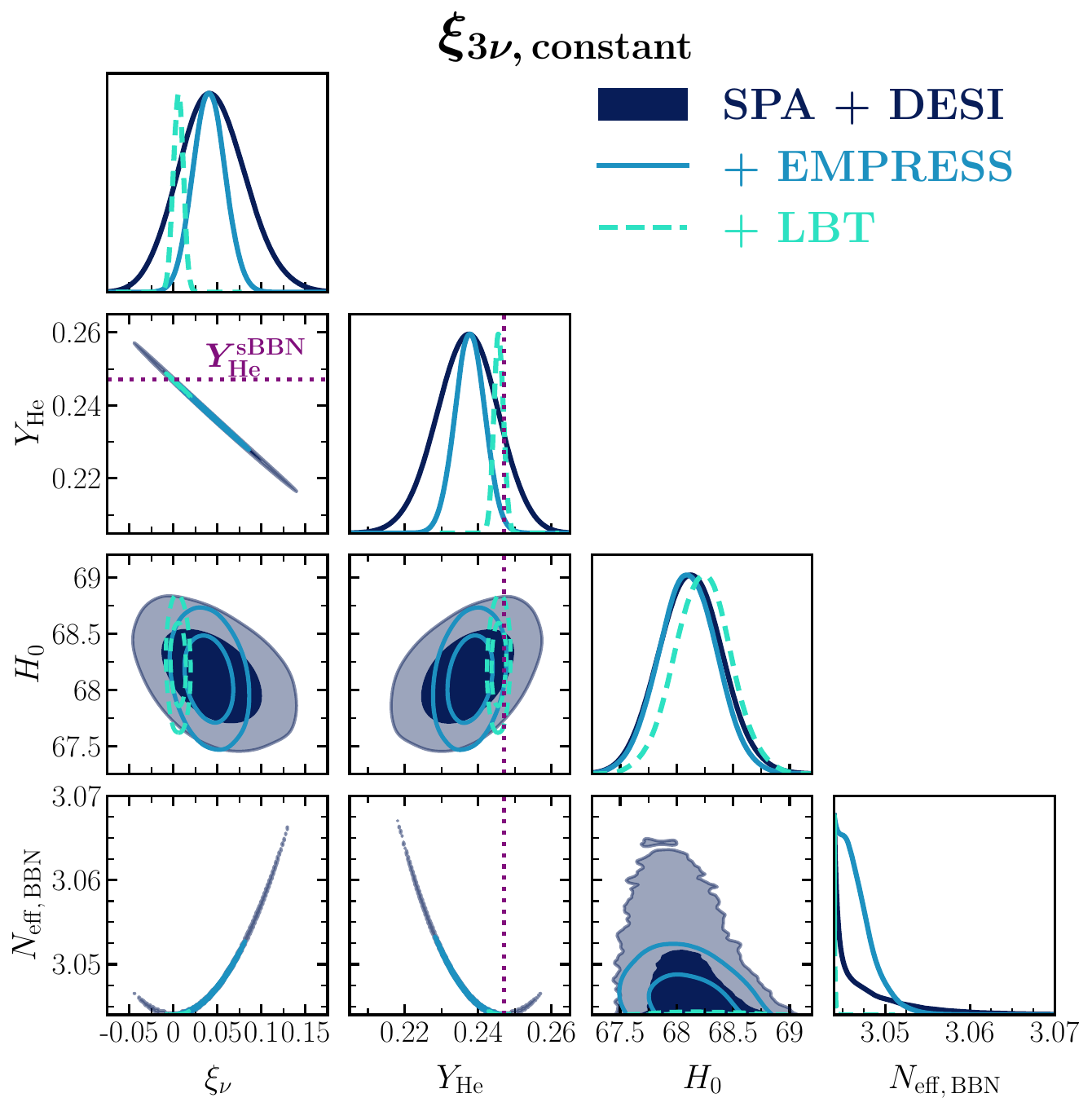}
 \caption{Triangular plots showing the 68 \% and 95 \% C.L. of a reduced set of parameters for the case of the three degenerate neutrinos and assuming the chemical potential to be constant. We additionally highlight the standard predicted value of the Helium fraction.} \label{fig:triangularcase3nuconstreduced}
\end{figure}

\Cref{fig:triangularcase3nuconstreduced} shows the marginalized posterior distributions for a reduced set of parameters sampled in the MCMC, in addition to the neutrino chemical potential.

When BBN measurements are not included, the contours enlarge, giving more freedom for $\xi_\nu$ to deviate from zero. It is interesting to notice that the posterior of $\xi_\nu$ is asymmetric: data prefer positive values over negative ones. This is a direct consequence of the effect that $\xi_\nu$ induces on $Y_\mathrm{He}$ during BBN; CMB data (and BBN data from EMPRESS) have a preference for slightly lower $Y_\mathrm{He}$ with respect to the LBT data on BBN, pushing $\xi_\nu$ to explore a region away from zero, see~\Cref{tab:case3}.

When BBN likelihoods are included in the analyses, whether based on EMPRESS or LBT measurements, the bounds tighten significantly and the limits on the Helium fraction converge toward the respective collaboration measurements. As expected, the tightest bounds are obtained when including the LBT measurement of the $Y_\mathrm{He}$, as it is currently the most precise determination of the primordial Helium fraction available.

It is interesting to note that, even though the EMPRESS measurement creates a mild tension with standard BBN predictions, the results obtained in that case are consistent with our baseline (\textbf{SPA + DESI}) scenario.

We have also computed the Bayesian evidence with respect to a $\Lambda$CDM model, see~\Cref{tab:case3}. Our results show that there is only weak evidence for the former model with respect to models with non-vanishing lepton asymmetries. There is only a moderate evidence for a $\Lambda$CDM Universe once LBT data are included in the analyses, which makes sense, given the fact that this dataset does not present any tension with the standard BBN prediction, and we find that cosmology alone prefers lower values for the Helium fraction, $Y_\mathrm{He}$. 

We finally highlight that, as expected, $N_\mathrm{eff, \, BBN}$  is a direct function of $\xi_\nu$, and follows the predicted parabola shape (see \Cref{eq:deltaneff}).

Note that we can only derive an upper bound on $N_\mathrm{eff, \, BBN}$ since we always work with a fixed value for $N_\mathrm{eff}^\mathrm{std}=3.044$, meaning that the impact of $\xi_\nu$ is only to increase its value\footnote{This comment remain true for all the considered analyses.}. Moreover, if we focus on the constraints presented in~\Cref{tab:case3} for $\Delta N_\mathrm{eff, \, BBN}$, we find them to be tighter than the ones presented by, e.g., the Planck collaboration. This is due to the fact that in our analysis, both $N_\mathrm{eff, \, BBN}$ and $\Delta N_\mathrm{eff, \, BBN}$ are derived parameters from $\xi_\nu$ (see~\Cref{eq:deltaneff}), characterized by a uniform prior. As a result, the induced prior on both parameters is non-uniform, assigning a larger prior volume to lower values of $\Delta N_\mathrm{eff, \, BBN}$.

\begin{figure}[h!]
    \centering
    \includegraphics[width=\linewidth]{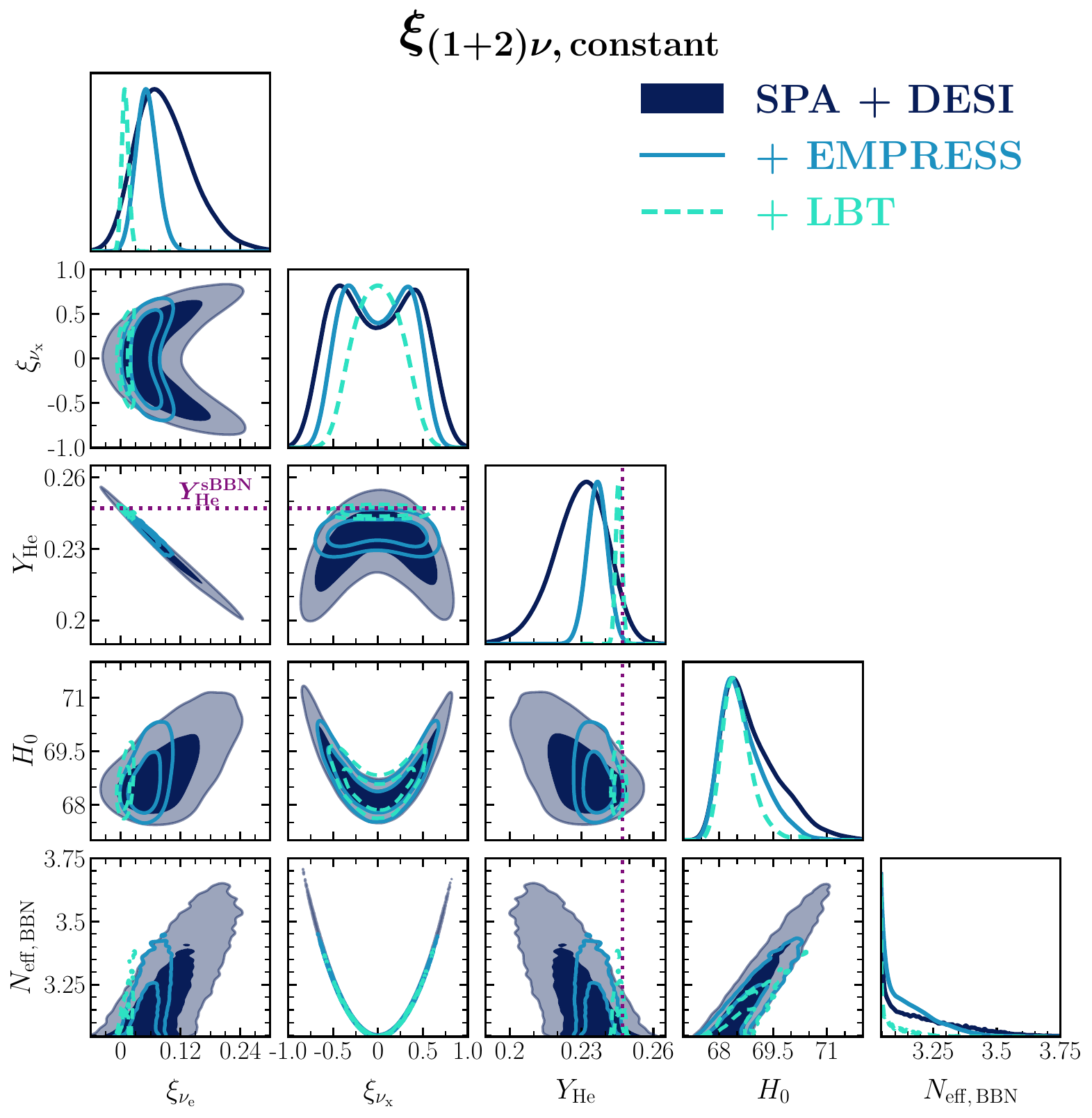}
    \caption{Triangular plots showing the 68 \% and 95 \% C.L. of a reduced set of parameters for the case in which we consider a (very) light neutrino state and the remaining two massive states to be degenerate. We still assume that the chemical potential of the neutrinos is constant. We additionally highlight the standard predicted value of the Helium fraction.}
    \label{fig:triangularcaseexnuconstreduced}
\end{figure}

\subsection{Case {\boldmath$\xi_{(1+2)\nu,\,\mathrm{constant}}$}}
\label{sec:12nuconst_results}

In \Cref{fig:triangularcaseexnuconstreduced} we show the marginalized posteriors for the same set of parameters as in the previous subsection, but distinguishing now between the lightest neutrino contribution and that of the massive species, treated as degenerate. This leads to splitting the previous $\xi_\nu$ into two parameters, namely $\xi_{\nu_\mathrm{e}}$ and $\xi_{\nu_\mathrm{x}}$, following \texttt{PArthENoPE} nomenclature.

The most interesting result is the bimodal posterior for $\xi_{\nu_\mathrm{x}}$, leading to ``banana-shaped'' contours in all panels involving this parameter, which is a direct consequence of the behaviour shown in \Cref{fig:parthenope}. Moreover, for the remaining of the effects on cosmology, the bimodal distribution is explained by the fact that in \Cref{eq:deltaneff,eq:omeganu} the sign of each $\xi_{\nu_\alpha}$ is physically irrelevant. As a consequence, the data cannot distinguish between positive and negative values of $\xi_{\nu_\mathrm{x}}$, but it can for $\xi_{\nu_\mathrm{e}}$ due to its important role in the weak interaction rates important for BBN.

Comparing \Cref{fig:triangularcaseexnuconstreduced} to \Cref{fig:triangularcase3nuconstreduced}, it is clear that, in the degenerate scenario, the bounds on $\xi_\nu$ are actually driven by the bounds on $\xi_{\nu_\mathrm{e}}$, while almost the whole range is allowed for $\xi_{\nu_\mathrm{x}}$, meaning that BBN (and cosmology more broadly) is mostly sensitive to $\xi_{\nu_\mathrm{e}}$. This is expected, as the electron neutrino is directly involved in BBN processes, specifically shifting the neutron–proton equilibrium ratio at freeze-out and thereby modifying the value of the primordial Helium abundance, whereas the other flavours mostly affect cosmology through their relativistic and non-relativistic energy densities. 

In comparison with the previous scenario (e.g., three degenerate massive neutrinos), there is no significant change regarding the impact of the different BBN likelihoods considered on the parameters, with a general tightening of the posteriors, more evident when we consider LBT. Moreover, the LBT likelihood has also the effect of suppressing the bimodal behavior of the $\xi_{\nu_\mathrm{x}}$ posterior, making it converge to a single peak centred in 0. This is due to the high precision of the LBT measurement and its agreement with the standard BBN value for the Helium fraction, effectively recovering the standard predictions and removing the degeneracies between the parameters. In contrast, the EMPRESS measurement prefers lower value of $Y_\mathrm{He}$, leaving more freedom for deviations of $\xi_{\nu_\mathrm{x}}$ from 0, thus not suppressing the bimodal distribution. It also allows more freedom to $\xi_{\nu_\mathrm{e}}$.

A notable feature of this scenario is the appearance of a tail in the posterior distribution of $H_0$, which relaxes its upper bound towards larger values, possibly alleviating the tension with Supernovae measurements~\cite{Riess:2021jrx}. We argue that this tail is due to the larger values of $\Neff$, compared to standard predictions, driven by the regions away from zero that $\xi_{\nu_\mathrm{e}}$ and, most significantly, $\xi_{\nu_\mathrm{x}}$ explore. This reflects the positive correlation between $H_0$ and $N_\mathrm{eff}$: a larger expansion rate in the early Universe reduces the sound horizon and keeping the angular acoustic scale constant requires an increase in the value of $H_0$. Notice that, compared to the previous scenario, the correlation between $\xi_{\nu_\mathrm{e}}$ and $H_0$ is positive, implying therefore a larger value of $H_0$, as the value of $\xi_{\nu_\mathrm{e}}>0$ (see~\Cref{tab:case12}).

We highlight that, in this scenario, $N_\mathrm{eff, \, BBN}$ is a function of both $\xi_{\nu_\mathrm{x}}$ and $\xi_{\nu_\mathrm{e}}$, and has a large tail towards higher values due to the wide region explored by $\xi_{\nu_\mathrm{x}}$. The correlation between the latter and $\xi_{\nu_\mathrm{e}}$, in which $\xi_{\nu_\mathrm{e}}$ is only varying within a small region is responsible for the narrowness of the parabola in the correlation with $N_\mathrm{eff, \, BBN}$.

Finally, considering the Bayesian evidence, we find a (non-signifcant) preference for our model over $\Lambda$CDM when we include the EMPRESS likelihood, otherwise $\Lambda$CDM is mildly preferred (see~\Cref{tab:case12}).

\begin{figure*}[t]
    \centering
    \includegraphics[width=\linewidth]{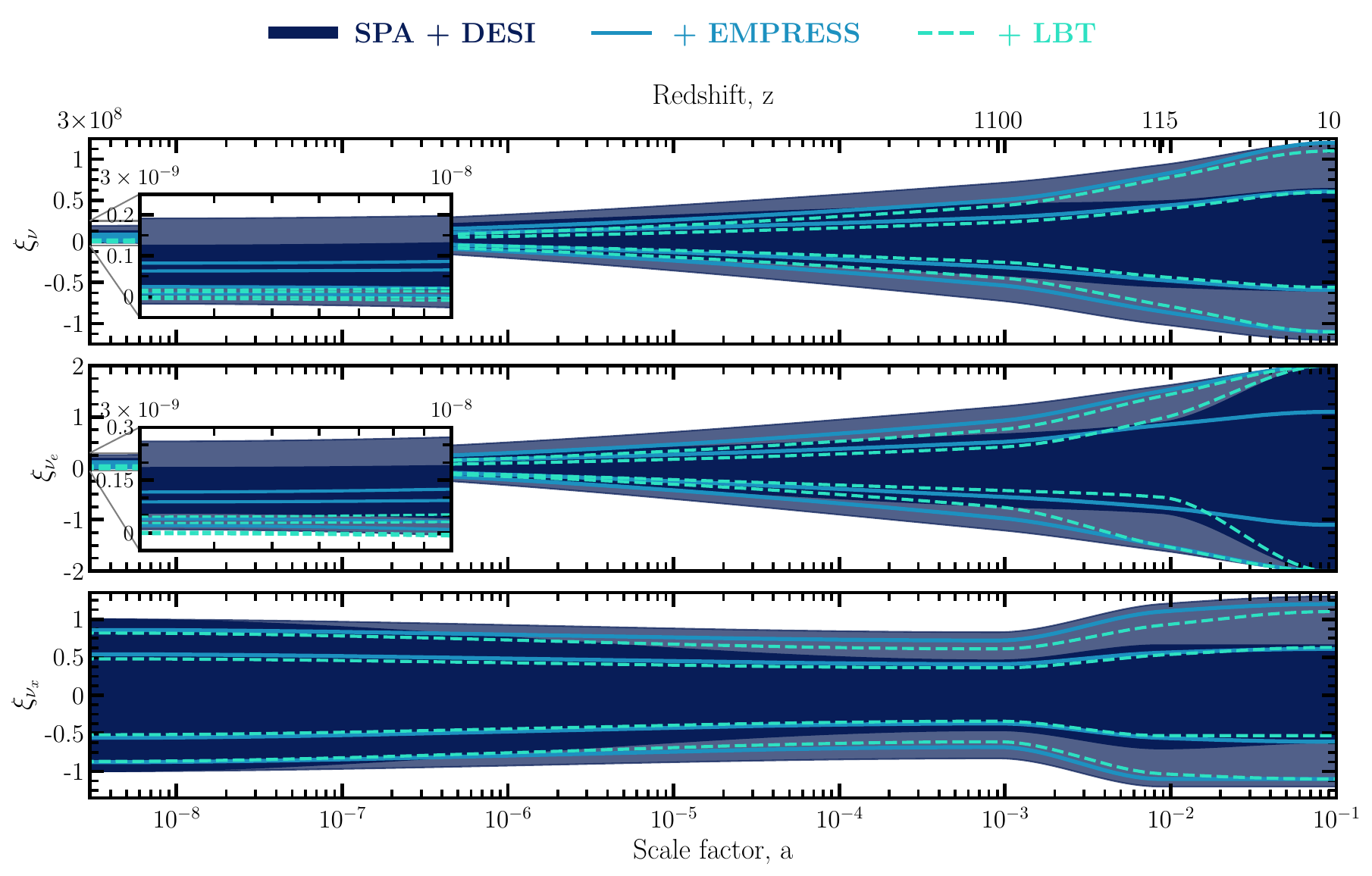}
    \caption{Time evolution of the neutrino chemical potentials, obtained using the PCHIP interpolation method. We show both the 68\% and 95\% C.L. bands. \textit{Top}: we show the chemical potential for the case of three degenerate neutrinos. \textit{Middle}: we show the chemical potential corresponding to the lightest state, denoted as $\xi_{\nu_\mathrm{e}}$. \textit{Bottom}: we show the common chemical potential corresponding to two massive degenerate neutrinos, denoted as $\xi_{\nu_\mathrm{x}}$. The colour code is like explained in the legend and consistent with the triangular plots.}
    \label{fig:reconstruction}
\end{figure*}

\begin{figure*}[t]
    \centering
    \includegraphics[width=\linewidth]{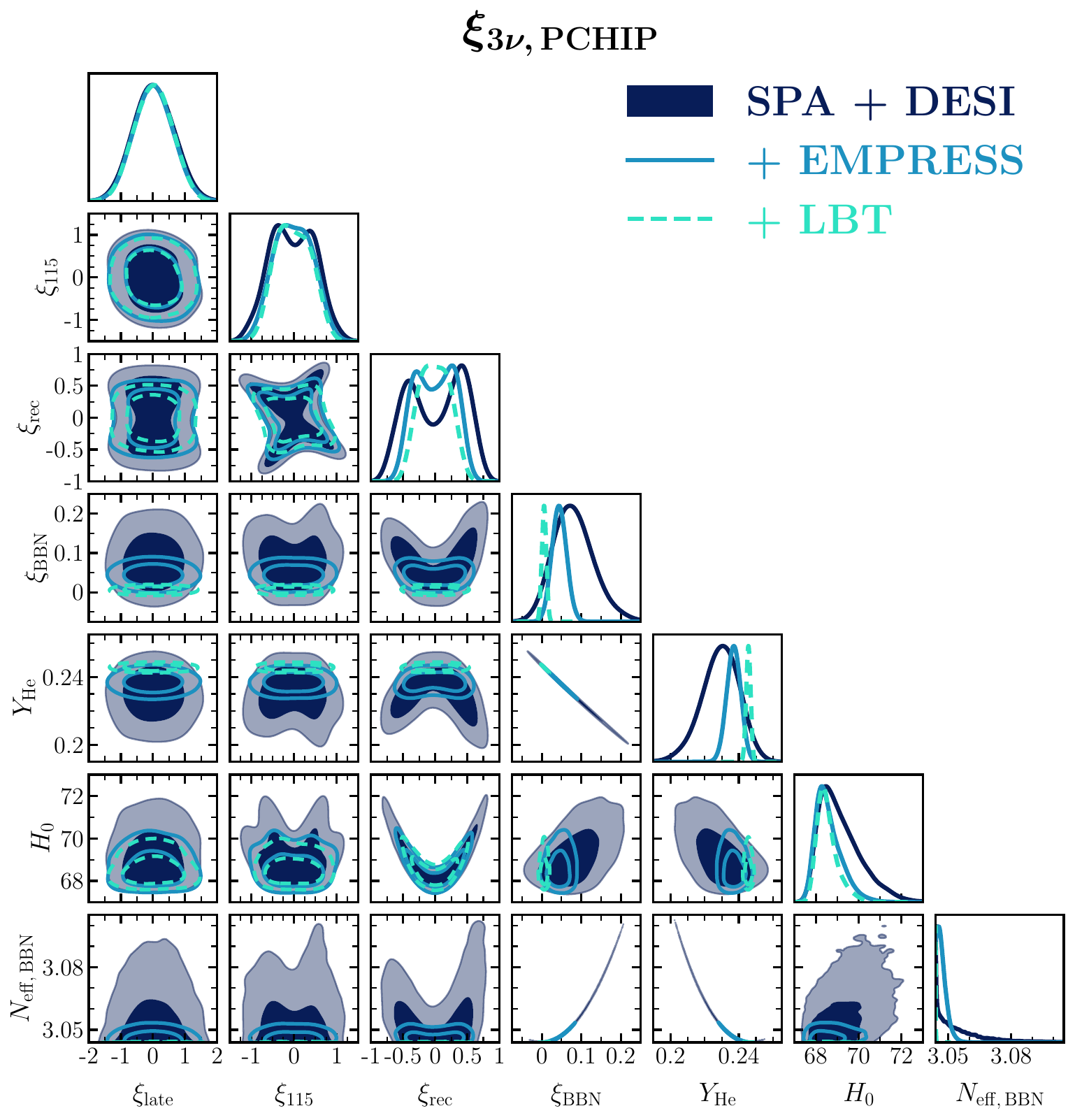}
    \caption{Triangular plots showing the 68 \% and 95 \% C.L. of a reduced set of parameters for the case of the three degenerate neutrinos and considering the chemical potential to be time-varying, assuming that the functional form is given by the PCHIP interpolating function.}
    \label{fig:triangularcase3nupchipreduced}
\end{figure*}

\begin{figure*}[t]
    \centering
    \includegraphics[width=\linewidth]{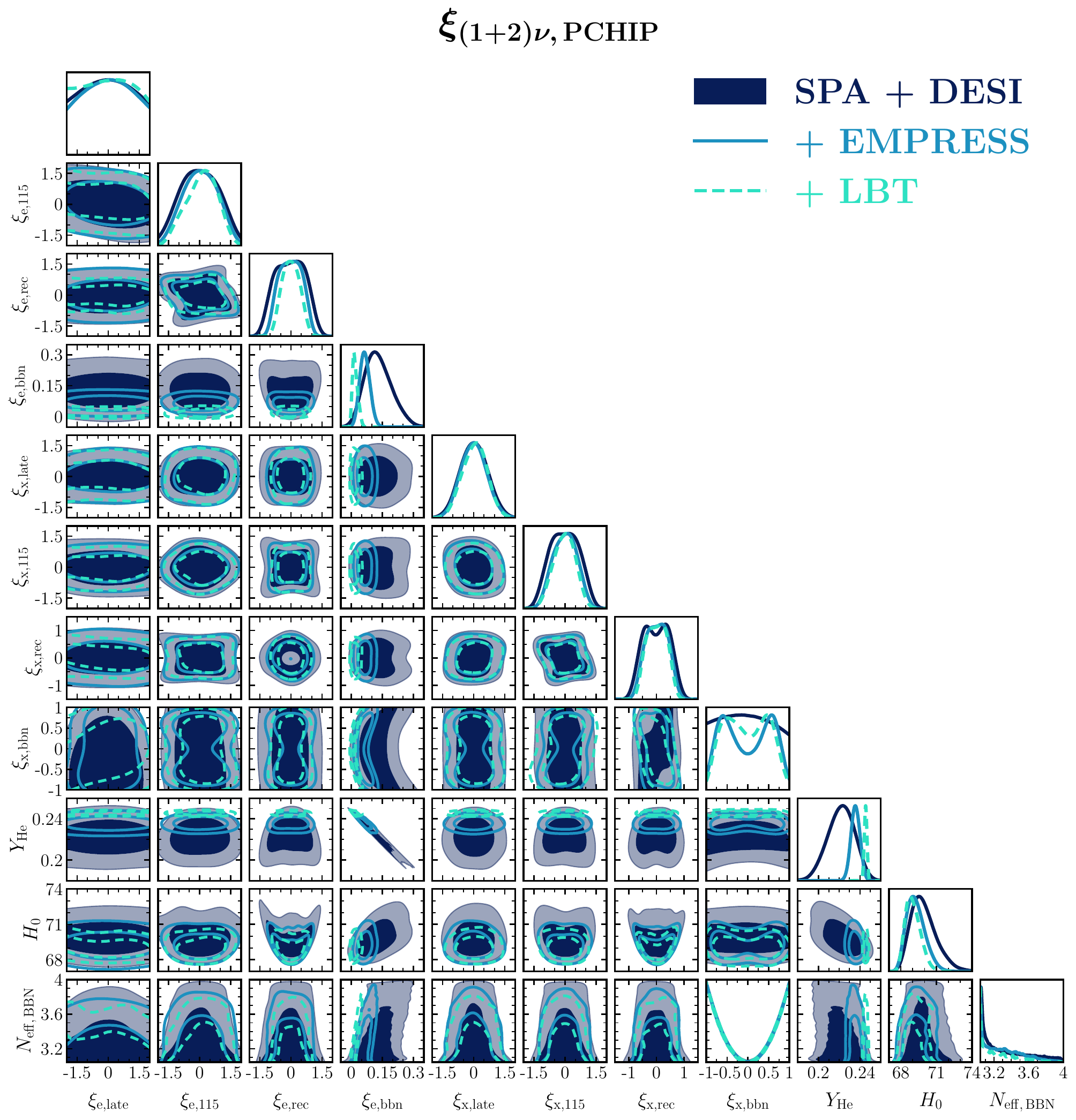}
    \caption{Triangular plots showing the 68 \% and 95 \% C.L. of a reduced set of parameters for the case in which we consider a (very) light neutrino state and the remaining two massive states to be degenerate. We present the case where the chemical potential is time-varying, assuming that the functional form is given by the PCHIP interpolating function.} 
    \label{fig:triangularcaseexnupchipreduced}
\end{figure*}

\subsection{Cases {\boldmath$\xi_{\nu, \, \mathrm{PCHIP}}$}}
\label{sec:pchip_results}

In \Cref{fig:reconstruction} we show the reconstruction of the neutrino chemical potential, obtained with the PCHIP method. We only show the evolution between the nodes listed in \Cref{tab:nodes}, since earlier than $z=3\times10^{-8}$ and later than $z=10$ the chemical potential has the same values as the respective limiting nodes (see~\Cref{sec:PCHIP}). We show both the 68\% and 95\% C.L. results, as dark-blue bands for our baseline data set and lines for the post-processing cases, respectively light-blue solid for including EMPRESS and cyan dashed for including LBT. In the top panel we show the reconstruction for the common $\xi_\nu$ of the $3\nu$ case, while in the middle and bottom panel we show the reconstructions for the two chemical potentials introduced in the $(1+2)\nu$ case, respectively $\xi_{\nu_\mathrm{e}}$ and $\xi_{\nu_\mathrm{x}}$.

\textbf{\emph{Case $\boldsymbol{3 \, \nu}$}$\,-\,$}As one would expect, we find that the most constraining epoch is around the BBN period (see~\Cref{fig:reconstruction}). This is due to the correlation between the chemical potential and the Helium fraction, to which CMB data are sensitive through the diffusion damping of photons, mostly affecting BBN predictions (see \Cref{fig:parthenope}). Adding BBN data makes the constraint significantly tighter at the BBN epoch, propagating the effect up to today, still leaving more freedom to the values that $\xi_\nu$ can take.

Interestingly, we find that the chemical potential is constrained, albeit not tightly, around recombination, when neutrinos hold about one-tenth of the total energy density of the Universe. We argue that this constraint derives from the massive nature of neutrinos, and the fact that variations in the chemical potential reflect onto $\Omega_\nu$, to which CMB spectra, in particular CMB lensing, are sensitive through e.g. the free-streaming. This propagates a bit at later times, as previously stated, albeit without adding a significant constraining power to set tight bounds on the parameter $\xi_\nu$. 

In \Cref{fig:triangularcase3nupchipreduced} we show the 68\% and 95\% C.L. triangle plot on the nodes used for the reconstruction and, also, on their correlations with $H_0$, $Y_\mathrm{He}$ and $N_\mathrm{eff, \, BBN}$.

Focusing on the nodes, we see the emergence of bimodal distributions for $\xi_{\nu, \, \mathrm{rec}}$ and $\xi_{\nu, \, 115}$, in a similar fashion to the bimodal distribution $\xi_{\nu_\mathrm{x}}$ (see \Cref{fig:triangularcase3nuconstreduced}). The reason for the bimodal posterior is an effect due to a mild preference for a larger $\Neff$ around recombination (see also Ref.~\cite{Wang_2025}), and for a slightly larger $\Omega_{\nu}$: such preferences are translated into values of $\xi_{\nu, \mathrm{rec}}$ and  $\xi_{\nu, \mathrm{115}}$ away from 0, see~\Cref{tab:case3}. The inclusion of the BBN measurements has the strongest impact on the BBN node, propagating lightly to the other nodes by killing the bimodal posterior distributions (particularly if we consider the LBT data, which drive $\xi$ to 0), while having zero effect on the late time node, consistently with our expectations.

Since we have more degeneracies between parameters and a time varying chemical potential, and therefore a time-varying $N_\mathrm{eff}$, we find a longer tail in the posterior distribution of $H_0$, partly lost once we include the BBN data. 

Finally, the results for the Bayesian evidence with respect to a $\Lambda$CDM model are shown in~\Cref{tab:case3}. In this case, our results show that there is only weak evidence for the PCHIP lepton asymmetry model with respect to the $\Lambda$CDM scenario when considering the basic dataset and also when adding EMPRESS observations to the baseline data. As in the constant case, there is only a mild moderate evidence for a $\Lambda$CDM Universe once LBT measurements are included in the analyses.

\textbf{\emph{Case $\boldsymbol{(1+2) \, \nu}$}$\,-\,$} 

It is particularly interesting how the effects driving the bounds - BBN and $\Neff$ in the early Universe, free-streaming through $\Omega_\nu$ in the late Universe - decouple, once we allow for a separation between $\xi_{\nu_\mathrm{e}}$ and $\xi_{\nu_\mathrm{x}}$.  During BBN, $\xi_{\nu_\mathrm{e}}$ directly modifies the neutron-to-proton conversion rates, and  is by far the most constrained quantity, with similar behaviour as the $\xi_\nu$ of the previous case, while $\xi_{\nu_\mathrm{x}}$, which only enters BBN indirectly through $\Neff$, is much less constrained by BBN light element abundances. Both quantities reach similar bounds around recombination, when the mass is still negligible. At later times, the very light neutrino state and the corresponding $\xi_{\nu_\mathrm{e}}$ which in practice does not contribute to $\Omega_\nu$, is much more loosely constrained with respect to $\xi_{\nu_\mathrm{x}}$. 

Adding BBN datasets, as one expects, modifies the profile of the $\xi_{\nu_\mathrm{e}}$ around BBN, pushing it towards slightly positive values, and the propagation of this modification to later epochs and to $\xi_{\nu_\mathrm{x}}$ is small. As a confirmation that the data is only sensitive to $\xi_{\nu_\mathrm{x}}$ through even powers, the contours of $\xi_{\nu_\mathrm{x}, \, i}$ are symmetrical at all epochs. 

In \Cref{fig:triangularcaseexnupchipreduced} we show the 68\% and 95\% C.L. triangle plot  on the nodes used for the reconstruction and, also, on their correlations with $H_0$, $Y_\mathrm{He}$ and $N_\mathrm{eff, \, BBN}$.

Focusing on the nodes, we see the emergence of bimodal distributions for the early $\xi_{\nu_\mathrm{x}}$ nodes, given the preference for larger integrated values, which are sensitive to even powers of the chemical potentials. However, the bimodal distribution does not appear in the first nodes of $\xi_{\nu_\mathrm{e}}$, due to its asymmetric effect in BBN physics. Adding BBN measurements has the same effect as in the previous case: it affects mostly the BBN epoch nodes, pushing $\xi_{\nu_e, \mathrm{BBN}}$ closer to the values measured by each collaboration. Curiously, before adding these measurements, the posterior for $\xi_{\nu_\mathrm{x}, \mathrm{BBN}}$ is not bimodal (compare with  \Cref{fig:triangularcaseexnuconstreduced}) perhaps due to a larger region explored by $\xi_{\nu_e, \mathrm{BBN}}$, allowing $\xi_{\nu_\mathrm{x}, \mathrm{BBN}}$ to remain closer to 0. When BBN is included, a bimodal appears for $\xi_{\nu_\mathrm{x}, \, \mathrm{BBN}}$ for both the datasets. While for EMPRESS the bimodal is a signal of a preference for higher $\Neff$, in the LBT case it might be due to not having properly explored the parameter region around $\xi_{\nu_\mathrm{x}, \mathrm{BBN}}$ = 0, as the contour of LBT in the $Y_\mathrm{He}$-$\xi_{\nu_\mathrm{x}, \mathrm{BBN}}$ lies at the 2$\sigma$ region. Particularly interesting, around recombination, we notice the appearance of a circumference in the plane of the two recombination nodes, $\xi_{\nu_\mathrm{e}, \, \mathrm{rec}}$ and $\xi_{\nu_\mathrm{x}, \, \mathrm{rec}}$, due to the level set of constant (and larger than 3.044, see~\Cref{tab:case12}) $\Neff$, expressed in terms of the two chemical potentials.

We find also in this case a longer tail in the posterior distribution of $H_0$, partly lost once we include the BBN data, a consequence of the tail on $N_\mathrm{eff, \, BBN}$. It is interesting to see that even when including LBT, which should recover the standard predictions, we still have a higher value of $H_0$, alleviating the $H_0$ tension. 

Finally, regarding the Bayes factors, we see moderate and strong evidences againts $\Lambda$CDM. We conjecture that this is due to the large number of parameters and to the time variation of some of them, able to accommodate possible discrepancies in the data, as the different $\Omega_\mathrm{m}$ preferred by BAO and CMB data, for example. As a matter of fact, this could provide a better fit to some features that otherwise are not accommodated by other analyses. Notice also that we are not reaching a level of convergence strong enough to actually reach significant conclusions on the results of this case and due to degeneracies between the parameters, we could be dealing with statistical features more than physical ones.

\begin{table*}[t]
\centering
\small\addtolength{\tabcolsep}{+3pt}
\def\arraystretch{1.4}
\begin{tabular}{cccc}
\multicolumn{4}{c}{\textbf{\emph{$\boldsymbol{\xi_{3\nu,\,\mathrm{constant}}}$}}} \\ \hline
 & \textbf{SPA + DESI} & \textbf{+ EMPRESS} & \textbf{+ LBT} \\
\hline
{\boldmath$100\theta_\mathrm{s}$} & $1.04170^{+0.00048}_{-0.00047}$ & $1.04171^{+0.00046}_{-0.00045}$ & $1.04178\pm0.00044$\\
{\boldmath$\tau_\mathrm{reio}$} & $0.066^{+0.012}_{-0.011}$ & $0.066^{+0.012}_{-0.011}$ & $0.066^{+0.012}_{-0.011}$\\
{\boldmath$\log(10^{10} A_\mathrm{s})$} & $3.064^{+0.023}_{-0.021}$ & $3.064^{+0.022}_{-0.020}$ & $3.067^{+0.022}_{-0.021}$ \\
{\boldmath$n_\mathrm{s}$} & $0.9692\pm0.0098$ & $0.9696\pm0.0068$ & $0.9735^{+0.0057}_{-0.0058}$\\
{\boldmath$\Omega_\mathrm{c} h^2$} & $0.1179\pm0.0012$ & $0.1179\pm0.0012$ & $0.1179\pm0.0012$\\
{\boldmath$\Omega_\mathrm{b} h^2$} & $0.02240\pm0.00025$ & $0.02238\pm0.00020$ & $0.02247\pm0.00018$\\
{\boldmath$\xi_\nu$} & $0.044^{+0.077}_{-0.071}$ & $0.041^{+0.036}_{-0.035}$ & $0.006\pm0.011$\\
{\boldmath$Y_\mathrm{He}$} & $0.237^{+0.016}_{-0.017}$ & $0.2378^{+0.0079}_{-0.0080}$ & $0.2456\pm0.0025$\\
{\boldmath$H_0$} & $68.13^{+0.56}_{-0.54}$ & $68.10^{+0.51}_{-0.50}$ & $68.23\pm0.49$\\
{\boldmath$\Delta N_\mathrm{eff, \, BBN}$} & $< 0.016$ & $< 0.007$ & $< 0.0003$\\
\hline
$\ln\mathcal{B}$ & $1.683\pm0.017$ & $0.285\pm0.026$ & $2.844\pm0.069$ \\
\hline
\hline
\multicolumn{4}{c}{\textbf{\emph{$\boldsymbol{\xi_{3\nu,\,\mathrm{PCHIP}}}$}}} \\ \hline
{\boldmath$100\theta_\mathrm{s}$} & $1.04126^{+0.00080}_{-0.00093}$ & $1.04152^{+0.00056}_{-0.00059}$ & $1.04168^{+0.00050}_{-0.00052}$\\
{\boldmath$\tau_\mathrm{reio}$} & $0.065^{+0.013}_{-0.011}$ & $0.066\pm0.012$ & $0.067^{+0.013}_{-0.012}$\\
{\boldmath$\log(10^{10} A_\mathrm{s})$} & $3.072^{+0.026}_{-0.024}$ & $3.070^{+0.025}_{-0.023}$ & $3.072\pm0.023$\\
{\boldmath$n_\mathrm{s}$} & $0.973^{+0.013}_{-0.011}$ & $0.9727^{+0.0096}_{-0.0091}$ & $0.9759^{+0.0081}_{-0.0075}$\\
{\boldmath$\Omega_\mathrm{c} h^2$} & $0.1213^{+0.0067}_{-0.0045}$ & $0.1195^{+0.0037}_{-0.0027}$ & $0.1189^{+0.0028}_{-0.0022}$\\
{\boldmath$\Omega_\mathrm{b} h^2$} & $0.02243\pm0.00025$ & $0.02242^{+0.00022}_{-0.00021}$ & $0.02250^{+0.00021}_{-0.00020}$\\
{\boldmath$\xi_{\nu, \,\mathrm{late}}$} & $0.0\pm1.2$ & $0.0^{+1.2}_{-1.1}$ & $0.0\pm1.1$\\
{\boldmath$\xi_{\nu , 115}$} & $-0.03^{+0.96}_{-0.98}$ & $-0.02\pm0.83$ & $-0.02^{+0.78}_{-0.75}$\\
{\boldmath$\xi_{\nu, \,\mathrm{rec}}$} & $0.02^{+0.69}_{-0.74}$ & $-0.01^{+0.051}_{-0.052}$ & $-0.01^{+0.44}_{-0.43}$\\
{\boldmath$\xi_{\nu, \,\mathrm{BBN}}$} & $0.081^{+0.11}_{-0.098}$ & $0.044^{+0.038}_{-0.036}$ & $0.006\pm0.011$\\
{\boldmath$Y_\mathrm{He}$} & $0.229^{+0.021}_{-0.022}$ & $0.2370^{+0.0081}_{-0.0082}$ & $0.2456\pm0.0026$\\
{\boldmath$H_0$} & $69.1^{+2.1}_{-1.5}$ & $68.8^{+1.5}_{-1.2}$ & $68.6^{+1.2}_{-0.88}$\\
{\boldmath$\Delta N_\mathrm{eff, \, BBN}$} & $< 0.04$ & $< 0.008$ & $< 0.0003$\\
\hline
$\ln\mathcal{B}$ & $-0.877\pm0.046$ & $-1.734\pm0.070$ & $2.086\pm0.053$ \\
\hline
\end{tabular}
\caption{95 \% C.L. bounds on the full set of parameters sampled in the MCMC for the case in which we consider three degenerate neutrinos, both for the case of assuming the chemical potential as a constant and as a function of time. We additionally present the bounds on the derived parameters $H_0$, $Y_\mathrm{He}$ and $\Delta N_\mathrm{eff, \, BBN}$, and the values of the Bayes factors.}
\label{tab:case3}
\end{table*}

\begin{table*}[t]
\centering
\small\addtolength{\tabcolsep}{+3pt}
\def\arraystretch{1.25}
\begin{tabular}{cccc}
\multicolumn{4}{c}{\textbf{\emph{$\boldsymbol{\xi_{(1+2)\nu,\,\mathrm{constant}}}$}}} \\ \hline
 & \textbf{SPA + DESI} & \textbf{+ EMPRESS} & \textbf{+ LBT} \\
\hline
{\boldmath$100\theta_\mathrm{s}$} & $1.04129^{+0.00080}_{-0.00092}$ & $1.04148^{+0.00058}_{-0.00061}$ & $1.04167^{+0.00049}_{-0.00055}$\\
{\boldmath$\tau_\mathrm{reio}$} & $0.065^{+0.012}_{-0.011}$ & $0.066^{+0.012}_{-0.011}$ & $0.067^{+0.012}_{-0.011}$\\
{\boldmath$\log(10^{10} A_\mathrm{s})$} & $3.067^{+0.023}_{-0.021}$ & $3.067^{+0.022}_{-0.021}$ & $3.069^{+0.021}_{-0.020}$\\
{\boldmath$n_\mathrm{s}$} & $0.971\pm0.010$ & $0.9725^{+0.0089}_{-0.0082}$ & $0.9752^{+0.0072}_{-0.0066}$\\
{\boldmath$\Omega_\mathrm{c} h^2$} & $0.1205^{+0.0053}_{-0.0037}$ & $0.1196^{+0.0036}_{-0.0027}$ & $0.1188^{+0.0025}_{-0.0021}$\\
{\boldmath$\Omega_\mathrm{b} h^2$} & $0.02242\pm0.00025$ & $0.02244^{+0.00024}_{-0.00021}$ & $0.02251^{+0.00020}_{-0.00019}$\\
{\boldmath$\xi_{\nu_\mathrm{e}}$} & $0.09^{+0.12}_{-0.11}$ & $0.052^{+0.045}_{-0.040}$ & $0.009^{+0.015}_{-0.013}$\\
{\boldmath$\xi_{\nu_\mathrm{x}}$} & $-0.02^{+0.73}_{-0.71}$ & $-0.01^{+0.58}_{-0.59}$ & $0.00^{+0.46}_{-0.45}$\\
{\boldmath$Y_\mathrm{He}$} & $0.229^{+0.021}_{-0.023}$ & $0.2365^{+0.0082}_{-0.0084}$ & $0.2455\pm0.0025$\\
{\boldmath$H_0$} & $68.9^{+1.7}_{-1.3}$ & $68.7^{+1.3}_{-1.0}$ & $68.52^{+0.90}_{-0.78}$\\
{\boldmath$\Delta N_\mathrm{eff, \, BBN}$} & $< 0.5$ & $< 0.4$ & $< 0.2$ \\
\hline
$\ln\mathcal{B}$ & $0.034\pm0.031$ & $-1.379\pm0.052$ & $2.707\pm0.033$ \\
\hline
\hline
\multicolumn{4}{c}{\textbf{\emph{$\boldsymbol{\xi_{(1+2)\nu,\,\mathrm{PCHIP}}}$}}} \\ \hline
{\boldmath$100\theta_\mathrm{s}$} & $1.04087^{+0.00093}_{-0.00098}$ & $1.04125^{+0.00065}_{-0.00068}$ & $1.04148^{+0.00058}_{-0.00061}$ \\
{\boldmath$\tau_\mathrm{reio}$} & $0.064\pm0.012$ & $0.066\pm0.012$ & $0.068\pm0.012$ \\
{\boldmath$\log(10^{10} A_\mathrm{s})$} & $3.076^{+0.030}_{-0.027}$ & $3.077^{+0.027}_{-0.025}$ & $3.079^{+0.024}_{-0.023}$ \\
{\boldmath$n_\mathrm{s}$} & $0.974^{+0.014}_{-0.012}$ & $0.976^{+0.011}_{-0.010}$ & $0.9784^{+0.0092}_{-0.0086}$  \\
{\boldmath$\Omega_\mathrm{c} h^2$} & $0.1238^{+0.0073}_{-0.0061}$ & $0.1215^{+0.0047}_{-0.0040}$ & $0.1203^{+0.0036}_{-0.0031}$  \\
{\boldmath$\Omega_\mathrm{b} h^2$} & $0.02244^{+0.00024}_{-0.00025}$ & $0.02248\pm0.00022$ & $0.02255^{+0.00020}_{-0.00018}$ \\
{\boldmath$\xi_{\nu_\mathrm{e},\,\mathrm{late}}$} & --- & --- & --- \\
{\boldmath$\xi_{\nu_\mathrm{e},\,115}$} & $0.0\pm1.6$ & $0.0\pm1.5$ & $0.1^{+1.3}_{-1.4}$ \\
{\boldmath$\xi_{\nu_\mathrm{e},\,\mathrm{rec}}$} & $0.0\pm1.2$ & $-0.02\pm0.94$ & $-0.01^{+0.76}_{-0.74}$ \\
{\boldmath$\xi_{\nu_\mathrm{e},\,\mathrm{BBN}}$} & $0.13^{+0.13}_{-0.12}$ & $0.066^{+0.050}_{-0.045}$ & $0.019^{+0.027}_{-0.022}$ \\
{\boldmath$\xi_{\nu_\mathrm{x},\,\mathrm{late}}$} & $0.0^{+1.3}_{-1.2}$ & $0.0^{+1.2}_{-1.1}$ & $0.0\pm1.1$ \\
{\boldmath$\xi_{\nu_\mathrm{x},\,115}$} & $0.0\pm1.2$ & $0.0\pm1.1$ & $-0.03^{+0.
95}_{-1.0}$ \\
{\boldmath$\xi_{\nu_\mathrm{x},\,\mathrm{rec}}$} & $0.00\pm0.83$ & $0.02\pm0.70$ & $0.01\pm0.60$ \\
{\boldmath$\xi_{\nu_\mathrm{x},\,\mathrm{BBN}}$} & --- & $-0.01\pm0.87$ & $-0.02^{+0.84}_{-0.85}$ \\
{\boldmath$Y_\mathrm{He}$} & $0.222^{+0.023}_{-0.026}$ & $0.2357^{+0.0083}_{-0.0084}$ & $0.2455^{+0.0025}_{-0.0026}$ \\
{\boldmath$H_0$} & $69.9^{+2.2}_{-1.9}$ & $69.3^{+1.6}_{-1.4}$ & $69.0^{+1.3}_{-1.1}$\\
{\boldmath$\Delta N_\mathrm{eff, \, BBN}$} & $< 0.9$ & $< 0.8$ & $< 0.7$\\
\hline
$\ln\mathcal{B}$ & $-4.80\pm0.13$ & $-5.62\pm0.16$ & $-2.85\pm0.16$ \\
\hline
\end{tabular}
\caption{95 \% C.L. bounds on the full set of parameters sampled in the MCMC for the case in which we consider the electron neutrino and the remaining two states to be degenerate, both for the case of assuming the chemical potential as a constant and as a function of time. We additionally present the bounds on the derived parameters $H_0$, $Y_\mathrm{He}$ and $\Delta N_\mathrm{eff, \, BBN}$, and the values of the Bayes factors.}
\label{tab:case12}
\end{table*}

\section{\label{sec:conclusions} Conclusions}

Contrarily to the case of the cosmic net baryon number, which is known to be small, there are no strong observational or theoretical constraints on the lepton number. The latter is a sum of contributions from the three neutrino flavours, although only the electron neutrino term is known to be small, as  $\xi_{\nu_\mathrm{e}} = \mu_{\nu_\mathrm{e}} / T$ is tightly constrained by Big Bang Nucleosynthesis (BBN). The degeneracy parameters of the other neutrino species, $\xi_{\nu_\mathrm{x}}$, remain poorly constrained by cosmological observations alone. While an increase in relativistic density during BBN due to non-zero neutrino chemical potentials will lead to an overproduction of Helium, a positive (negative) electron neutrino chemical potential directly modifies the weak interaction rates, reducing (increasing) the neutron-to-proton ratio and consequently decreasing (increasing) the primordial Helium abundance.

The total lepton number could be large, or it could be comparable to the baryonic one. If it is comparable to the baryon number, this could occur because all neutrino terms were small, but it could also be due to cancellations. Even if in the very early Universe the neutrino chemical potentials are tiny, non-zero neutrino chemical potentials could arise at later times due to particle decays (or other mechanisms). Therefore, it is mandatory not only to consider separately the electron neutrino chemical potential constraints from the other two flavours, but also to consider the possibility of time-dependent neutrino  degeneracy parameters. Motivated by these possibilities, in this work we compute up-to-date bounds on $\xi_{\nu_\mathrm{e}}$ and $\xi_{\nu_\mathrm{x}}$ assuming that they are either constant free-parameters along the cosmic history or that they are redshift dependent quantities. For the redshift dependent case, we base our analyses on the  Piecewise Cubic Hermite Interpolating Polynomial (PCHIP) method, using four nodes in redshift (around $z\simeq$ 10, 100, 1000 and $10^8$). 

By combining the latest CMB data from Planck, SPT, and ACT with DESI BAO measurements and complementary information from BBN observables, either from the EMPRESS or the LBT measurements, our analysis demonstrate that BBN observables are the most powerful probes of a possible non-zero neutrino chemical potential. As a matter of fact, it is only when we include them that we are able to place tighter constraints on the parameters of interest.

Our analyses explicitly show that the BBN data, via the change in neutron-to-proton interconversion rates, mostly constrain $\xi_{\nu_\mathrm{e}}$, parameter for which we observe a preferred non-zero positive value at $95\%$~C.L. in the non-degenerate neutrino case at the BBN period. The corresponding constraints on $\xi_{\nu_\mathrm{x}}$ remain instead comparatively weak. Any lepton asymmetry generated after BBN remains weakly constrained by current cosmological data within the phenomenological framework we considered in this work, which remains agnostic on the microphysical mechanism for the generation of such a lepton asymmetry.  Nevertheless, it is still true that from all the observations considered in the analyses, we have proved that the BBN period is the time providing the tightest constraints on $\xi_{\nu}$. 

Since the Hubble constant is correlated with $\xi_{\nu}$, through $N_\mathrm{eff}$, a larger value of $H_0$ is naturally reached within these models, making them very attractive schemes to alleviate the long standing $H_0$ tension, while remaining compatible with current cosmological observations.

Finally, our Bayesian model comparison analysis does not show any strong statistical preference for the minimal $\Lambda$CDM scenario over models where we could allow for (non-zero) lepton asymmetries; only a (mild) moderate preference appears when including LBT observations in the data fits. Future (and more precise) BBN measurements have the key to further constrain lepton asymmetries in the early Universe.

\begin{acknowledgments}
We thank Valerie Domcke, Julien Froustey and Nicola Barbieri for useful discussions about the project.

This article is based upon work from the COST Action CA21136 - “Addressing observational tensions in cosmology with systematics and fundamental physics (CosmoVerse)”, supported by COST - “European Cooperation in Science and Technology”. 
P.G.\ is supported by the SO project CEX2023-001292-S funded by MCIU/AEI/10.13039/501100011033 and has received funding from the European Union’s Horizon Europe programme under Marie Skłodowska-Curie Actions – Staff Exchanges (SE) grant agreement No 101086085 -ASYMMETRY. R.I.\ acknowledges support from the COST Action COSMIC WISPers CA21106, supported by COST. This work has also been supported by the Spanish grant PID2023-148162NB-C22. S.G.\ is supported by the Research grant TAsP (Theoretical Astroparticle Physics) funded by Istituto Nazionale di Fisica Nucleare (INFN), through the Ram\'on y Cajal contract RYC2023-044611-I funded by MICIU/AEI/10.13039/501100011033 and FSE+, and by the Spanish grant PID2023-147306NB-I00 (MCIU/AEI/10.13039/501100011033). D.W.\ thanks the support from the CDEIGENT fellowship of the Consejo Superior de Investigaciones Científicas (CSIC).

P.G.\ would like to thank CERN and the University of Sheffield for their hospitality during the completion of this work. 
\end{acknowledgments}

\bibliographystyle{apsrev4-2}
\bibliography{apssamp}

\end{document}